\documentclass[10pt]{article}

\usepackage[T1]{fontenc}
\usepackage[utf8]{inputenc}

\usepackage[margin=1in]{geometry}
\usepackage{url}
\usepackage{hyperref}
\usepackage{setspace}
\usepackage{lmodern}

\usepackage{cite}
\usepackage{booktabs}
\usepackage{tikz}
\usetikzlibrary{arrows.meta, positioning, shapes.geometric}
\usepackage{listings}
\begin{document}

\begin{center}
    {\LARGE Explainability and Certification of AI-Generated Educational Assessments}\\[8pt]
    Antoun Yaacoub$^{1}$, Zainab Assaghir$^{2}$, Anuradha Kar$^{1}$\\[4pt]
    $^{1}$aivancity School for Technology, Business \& Society, 151 Bd Maxime Gorki, 94800 Villejuif, France\\
    $^{2}$Faculty of Science, Lebanese University, Hadath Campus, Hadath, Lebanon\\[4pt]
    yaacoub@aivancity.ai, zainab.assaghir@ul.edu.lb, kar@aivancity.ai
\end{center}

\noindent\textbf{Abstract} \\
The rapid adoption of generative artificial intelligence (AI) in educational assessment has created new opportunities for scalable item creation, personalized feedback, and efficient formative evaluation. However, despite advances in taxonomy alignment and automated question generation, the absence of transparent, explainable, and certifiable mechanisms limits institutional and accreditation-level acceptance. This chapter proposes a comprehensive framework for explainability and certification of AI-generated assessment items, combining self-rationalization, attribution-based analysis, and post-hoc verification to produce interpretable cognitive-alignment evidence grounded in Bloom's and SOLO taxonomies. A structured certification metadata schema is introduced to capture provenance, alignment predictions, reviewer actions, and ethical indicators, enabling audit-ready documentation consistent with emerging governance requirements. A traffic-light certification workflow operationalizes these signals by distinguishing auto-certifiable items from those requiring human review or rejection. A proof-of-concept study on 500 AI-generated computer science questions demonstrates the framework's feasibility, showing improved transparency, reduced instructor workload, and enhanced auditability. The chapter concludes by outlining ethical implications, policy considerations, and directions for future research, positioning explainability and certification as essential components of trustworthy, accreditation-ready AI assessment systems.

\vspace{6pt}

\section{Introduction}

Generative AI has emerged as a transformative force in educational assessment, enabling rapid creation of multiple-choice questions (MCQs), distractors, worked examples, and formative feedback. Large language models (LLMs) can generate substantial quantities of pedagogically relevant items at a fraction of the time required for manual development, offering institutions new opportunities to enhance scalability and support curriculum design. Despite these advances, the integration of AI-generated assessments into accredited educational environments remains limited by concerns surrounding validity, transparency, fairness, and long-term reproducibility. Educators and quality assurance bodies increasingly question how AI-produced items are generated, whether they align with intended learning outcomes, and how institutions can document and defend their use during accreditation cycles.

Research in cognitive alignment and AI-generated questions has demonstrated that generative systems can approximate Bloom's or SOLO taxonomy levels through linguistic and structural cues \cite{yaacoub2025_alignment,yaacoub2025_solo}. Additional work has analysed readability, challenge levels, and feedback characteristics of AI-generated content, highlighting both strengths and limitations of current models \cite{yaacoub2025_feedback,yaacoub2025ijcnn}. However, these efforts do not yet offer a unified framework that connects generative outputs with explicit, explainable rationales or certification processes suitable for institutional governance.

The need for transparency has become more urgent as global AI regulations, including the EU Artificial Intelligence Act and UNESCO's guidelines on AI in education, increasingly classify educational assessment as a high-risk domain requiring documented oversight, data provenance, and human review \cite{eu_ai_act,unesco2023}. Traditional assessment workflows rely on established quality assurance mechanisms, but AI-generated items introduce novel challenges: they may embed hidden biases, exhibit inconsistent cognitive depth, or produce explanations that are not pedagogically interpretable. Consequently, institutions require mechanisms that bridge the gap between generative capabilities and accreditation expectations.

This chapter addresses these challenges by introducing a framework that integrates explainability, metadata-driven certification, and audit workflows into a cohesive process for managing AI-generated assessments. Explainability is approached through complementary methods---including self-rationalization, attribution analysis, and independent verification---to ensure that cognitive alignment is both detectable and interpretable. A structured certification metadata schema operationalizes these outputs into audit-ready documentation capturing provenance, alignment evidence, reviewer notes, and ethical risk indicators. Building on this schema, a traffic-light certification model distinguishes high-confidence items from those requiring human oversight or rejection. Finally, a proof-of-concept implementation using 500 AI-generated computer science questions demonstrates how this framework can enhance transparency, reduce instructor workload, and support accreditation compliance.

By embedding explainability and certification into the assessment generation pipeline, this chapter contributes a governance-ready approach to deploying generative AI in higher education. The proposed framework supports not only effective and scalable question generation but also ethical, transparent, and institutionally defensible assessment practices.

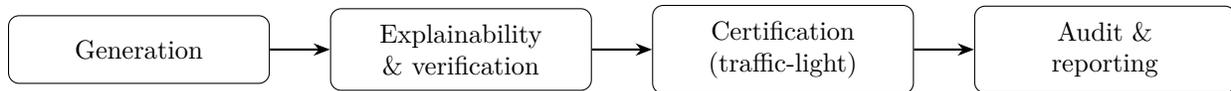
\begin{figure}[t]
\centering
\begin{tikzpicture}[
  box/.style={draw, rounded corners, align=center, inner sep=6pt, minimum height=9mm, minimum width=0.21\linewidth},
  arrow/.style={-{Stealth[length=2.2mm]}, thick},
  node distance=8mm
]
\node[box] (gen) {Generation};
\node[box, right=8mm of gen] (xai) {Explainability\\ \& verification};
\node[box, right=8mm of xai] (cert) {Certification\\ (traffic-light)};
\node[box, right=8mm of cert] (audit) {Audit \&\\ reporting};
\draw[arrow] (gen) -- (xai);
\draw[arrow] (xai) -- (cert);
\draw[arrow] (cert) -- (audit);
\end{tikzpicture}
\caption{Visual overview of the proposed pipeline: generation $\rightarrow$ explainability $\rightarrow$ certification $\rightarrow$ audit and reporting.}
\label{fig:pipeline}
\end{figure}

Figure~\ref{fig:pipeline} provides a concise visual summary of the proposed end-to-end process, from item generation to audit-ready reporting.

\section{Background and Literature Review}

\subsection{AI in Educational Assessment}

The domain of educational assessment has witnessed significant transformation as AI technologies progress from supporting scoring and item-delivery to enabling generative assessment design and response evaluation. Early systems emphasized computer-based testing and adaptive feedback  \cite{shute2013stealth,baker2014edm}. With the advent of large language models (LLMs) and generative AI (GenAI), new possibilities emerge: automatic generation of multiple-choice questions (MCQs), distractors, domain-specific worked examples, and automated rubrics \cite{yaacoub2024_oneclick,zawacki2019systematic}.

Contemporary studies indicate both opportunity and risk. A 2025 study published in Information outlines a "framework for generative AI-driven assessment in higher education," remarking that while generative tools enable rapid item creation and adaptive evaluation, they often lack pedagogical framing, transparency of generation processes, and robust quality assurance protocols \cite{ilieva2025framework}. A scoping review of generative AI in higher education assessment found primarily research from 2023 onward and identified a need for institutional-level policies and digital literacy among educators \cite{chiu2024ailiteracy}.

Moreover, studies of LLM-based question generation and classification indicate that while models can produce linguistically coherent items, there are substantive challenges in calibrating cognitive demand, distractor plausibility, and domain-validity. A recent systematic review pointed out that empirical evidence remains limited regarding the validity and reliability of AI-generated assessments in operational institutional contexts \cite{zhao2024}.

In parallel, empirical investigations have started to quantify the quality of AI-generated MCQs by examining taxonomy alignment, linguistic features and feedback properties, highlighting both the promise and the limitations of current generative approaches in real courses \cite{yaacoub2025_alignment,yaacoub2025_solo,yaacoub2025_feedback}.

These developments underscore a central tension: generative AI can scale item production and personalise assessment, yet it introduces issues of pedagogical alignment, transparency, validity, and institutional auditability that traditional assessment frameworks were not designed to handle.

\subsection{Cognitive Frameworks in Assessment Design}

Cognitive taxonomies remain foundational to aligning assessments with intended learning outcomes. The widely-used framework by Anderson \& Krathwohl \cite{anderson2001} reorganises Bloom's original taxonomy into six cognitive processes (Remember, Understand, Apply, Analyse, Evaluate, Create). SOLO Taxonomy \cite{biggs2014solo} provides another axis: the structural complexity of student responses from unistructural to extended abstract.

In contemporary practice, automated methods for taxonomy alignment have begun to surface. For instance, classifiers using linguistic features and action-verbs attempt to predict Bloom level; structural complexity and discourse markers have been used to estimate SOLO level \cite{li2022pafpn}. However, these systems often output a label or scalar score without human-interpretable rationale.

	More recent work explores how AI-generated questions themselves can be evaluated for taxonomy alignment. For example, \cite{ebner2025} introduced the QUEST framework which assesses LLM-generated MCQs across five dimensions: Quality, Uniqueness, Effort, Structure, and Transparency. Other conceptual studies emphasise the development of evaluative judgment in learners when generative AI contributes to both item generation and student work. These contributions illustrate that taxonomy alignment, while increasingly automated, still lacks the depth of interpretability and justification required by formal educational measurement and accrediting bodies.

Complementing these developments, a series of studies has shown that Bloom- and SOLO-level alignment of AI-generated questions can be automatically assessed with reasonable accuracy, using action-verb patterns, semantic similarity and lightweight classifiers, across multiple cohorts and courses \cite{yaacoub2025_alignment}, \cite{yaacoub2025_solo}. Additional work examining linguistic properties, readability, and challenge levels of AI-generated feedback further highlights the interplay between lexical characteristics and cognitive depth \cite{yaacoub2025_feedback}. These results provide an empirical foundation for integrating cognitive taxonomies into generative assessment pipelines, but they also highlight the absence of explicit, human-interpretable rationales accompanying alignment decisions.

\subsection{Explainable AI in Educational Systems}

Explainable AI (XAI) aims to provide insight into how AI systems produce their outputs, supporting transparency, trust, accountability and auditability \cite{doshi2017,guidotti2018}. Techniques such as LIME \cite{ribeiro2016}  and SHAP \cite{lundberg2017} highlight feature-importance or token-attribution, while attention or gradient-based saliency visualizations reveal internal model behaviour.

XAI is increasingly essential in education, where decisions affect learners' progression, fairness, and evaluation outcomes \cite{ifenthaler2020,holstein2019}. Research in learning analytics shows that transparency fosters trust and improves instructor willingness to adopt AI-based systems. However, most existing educational XAI work focuses on analytics dashboards, learner models, or predictive risk scoring, not on generative assessment technologies.

These works point to the growing feasibility of embedding explainability in assessment-generation systems. Yet, the literature also emphasises that explanations must be traceable (linked to specific features/concepts), consistent (repeatable across items), pedagogically meaningful (connected to taxonomy and learning outcomes), and auditable (recorded for review).
Recent empirical studies in educational AI have explored prompt-based self-rationalisation strategies, where the model is instructed to generate both an item and an explanation of its intended cognitive level, alongside post-hoc classification signals and linguistic markers; this combination enables a richer inspection of how questions relate to Bloom and SOLO categories \cite{yaacoub2025_alignment}. However, these explanations are typically stored informally and are not yet integrated into structured metadata schemas suitable for institutional audit.

The gap remains in transforming explanation outputs into institutional-grade certification metadata.

\subsection{Accreditation, Certification, and Governance Requirements}

Institutional accreditation and educational measurement standards increasingly recognise the need for transparent, valid, and reliable assessment practices. Bodies such as the Council for Higher Education Accreditation (CHEA) emphasise alignment of assessment with learning outcomes, fairness, validity, and documentation of processes. However, the introduction of generative AI into assessment complicates these requirements \cite{chea2022}.

Regulatory documents now address the role of GenAI in assessment. For example, university-level guidance for the use of generative AI in assessment emphasises human-in-the-loop oversight, transparency, privacy and accountability. In the domain of educational measurement, recent work explores ethical concerns such as algorithmic bias, automation bias, environmental cost and transparency in AI-enabled measurement.

From a governance perspective, institutions must integrate metadata around model version, prompt history, rationales for alignment and human review logs, to satisfy audit requirements. Sector reports on generative AI in higher education assessment argue that detection tools are often unreliable, and thus institutions should shift toward redesigning assessment types and developing explicit governance mechanisms rather than relying solely on plagiarism or AI-use detection.
Existing studies provide valuable insights into automated taxonomy alignment, linguistic indicators, and quality analysis of AI-generated questions. However, none of these approaches provide a formal certification mechanism or a governance-ready metadata structure. Current systems do not record provenance information, generation parameters, alignment rationales, or human review logs in a standardized way. As a result, institutions lack the means to trace how an item was generated, why a specific cognitive level was assigned, or how reviewers validated it over time. No comprehensive frameworks currently connect explainability outputs with accreditation or audit requirements, leaving a significant gap between technical evaluation and institutional governance.

These governance imperatives highlight that generative AI assessment systems cannot simply be "plugged in"; they require structured certification metadata, audit trails, explainable processes and institutional policy alignment.

\subsection{Synthesis and Research Gap}

The literature reviewed indicates the following:
\begin{itemize}
    \item Generative AI technologies are rapidly reshaping assessment design and administration, enabling scale, adaptability and new feedback mechanisms---but they introduce risks regarding validity, transparency, integrity and fairness.
    \item Automated cognitive taxonomy alignment techniques exist but frequently lack interpretable justification or pedagogical explanation.
    \item Explainability methods from XAI literature are beginning to be applied in educational assessment contexts, yet they are not yet integrated into operational assessment pipelines with audit-ready metadata.
    \item  Accreditation, certification and governance frameworks impose demands for transparency, traceability and human oversight that current generative assessment systems typically do not satisfy.
 
\end{itemize}

\subsubsection{Comparison with Existing Approaches}

To further clarify the chapter's contribution, Table~\ref{tab:comparison} contrasts the proposed framework with representative strands of related work. Prior approaches typically provide either cognitive-alignment signals, quality guidelines, or stand-alone explainability techniques, but they rarely combine these elements with a formal certification decision process and audit-ready documentation.

\begin{table}[t]
\centering
\caption{Concise comparison between the proposed framework and representative existing approaches for AI-generated assessment.}
\label{tab:comparison}
\begin{tabular}{@{}p{0.30\linewidth}p{0.14\linewidth}p{0.14\linewidth}p{0.14\linewidth}p{0.18\linewidth}@{}}
\toprule
Approach & Explainability & Certification & Auditability & Typical output \\ \midrule
Taxonomy-alignment classifiers / metrics \cite{yaacoub2025_alignment,yaacoub2025_solo} & Limited & No & No & Labels / confidence \\
MCQ quality frameworks (e.g., QUEST) \cite{ebner2025} & Partial & No & No & Heuristics / rubrics \\
Generic XAI methods (LIME/SHAP; surveys) \cite{ribeiro2016,lundberg2017,guidotti2018} & Yes & No & No & Attributions / insights \\
Proposed framework (this chapter) & Yes & Yes & Yes & Certified item + metadata \\ \bottomrule
\end{tabular}
\end{table}

Empirical studies on AI-generated MCQs, taxonomy alignment, and feedback analysis provide important building blocks toward addressing these challenges, showing how Bloom and SOLO alignment, linguistic indicators and feedback structure can be quantified and used as quality signals at scale \cite{yaacoub2025_alignment,yaacoub2025_solo,yaacoub2025_feedback}. However, these contributions are not yet embedded within a full governance and accreditation workflow.

Therefore, there remains a significant gap: a unified framework that integrates generative assessment item creation, cognitive alignment with taxonomy, automated explainability, and structured certification metadata---within an audit-friendly workflow that satisfies accreditation and governance requirements.

The remainder of the chapter addresses this gap by proposing a cohesive architecture for explainable, certifiable, accreditation-ready AI-generated assessments, underpinned by a metadata schema, traffic-light certification label, and audit workflow.

\section{Explainability in Assessment Generation}

Explainability in assessment generation concerns making the processes by which AI systems create, classify, and validate assessment items transparent to instructors, students, and accreditation bodies. In the context of AI-generated MCQs, explainability must clarify why a question targets a specific cognitive level, how its distractors were derived, and which aspects of the item support its pedagogical intent. Recent work in explainable educational assessment and LLM-as-judge systems has shown that natural language rationales, attribution maps, and calibrated verification models can substantially improve trust and usability, but these techniques are still rarely integrated into full assessment-generation pipelines. 

The framework adopted in this chapter organizes explainability into three complementary layers---self-rationalization, attribution-based methods, and post-hoc verification---whose outputs are then mapped to pedagogical constructs such as Bloom's and SOLO taxonomies. Together, these layers provide multiple, partly redundant perspectives on each generated item, enabling both technical scrutiny and educational interpretation.

\subsection{Self-Rationalization}

Self-rationalization denotes the capacity of LLMs to generate free-text explanations alongside their primary outputs. In assessment generation, this means that each MCQ can be accompanied by a rationale describing (i) why the stem and options correspond to a particular taxonomy level and (ii) why the correct answer is preferred over distractors. Self-rationalizing LLMs have been shown to improve judgment quality in generic evaluation tasks, where models act as fine-grained judges and produce chain-of-thought style justifications for their decisions. 

In educational contexts, self-rationalization has been applied to automated scoring and feedback. Studies on automatic scoring of science responses and essay scoring frameworks such as QwenScore+ show that rubric-aware prompts and chain-of-thought explanations can produce richer, more transparent feedback aligned with human scoring criteria. 

Similarly, AERA Chat demonstrates an LLM-based system that provides both scores and natural language rationales for student responses, allowing educators to visually inspect and critique the model's reasoning. 

Applied to AI-generated MCQs, self-rationalization can be operationalised in two main ways:

\begin{itemize}
    \item  Question-level rationales: when generating an item, the model is prompted to state the intended taxonomy level (e.g., Bloom "Apply" or SOLO "Relational") and to justify this choice in terms of required cognitive operations (e.g., recall vs. comparison vs. generalisation).
    \item  	Option-level rationales: the model explains why the correct option satisfies the stem and why each distractor is plausible yet incorrect.

\end{itemize}

Prompt engineering plays a central role in obtaining stable rationales. Work on lightweight prompt engineering for cognitive alignment in educational AI shows that explicit, detailed prompts specifying cognitive level and expectations lead to more accurate alignment than simpler or persona-based prompts \cite{yaacoub2025_prompt}. 

These findings can be extended by instructing the model not only to target a level, but also to articulate the reasoning that supports this target.

Self-rationalization alone, however, does not guarantee reliability. LLM explanations can be fluent yet partially incorrect or inconsistent with underlying decisions. Consequently, rationales must be complemented with more systematic, model-agnostic forms of explanation and verification.

\subsection{Attribution-Based Methods}
Attribution-based methods provide a second layer of explainability by quantifying which parts of the input contributed most to a model's decision. Classical techniques such as LIME and SHAP approximate the local decision boundary of a classifier by perturbing inputs and measuring the impact on predictions, yielding per-feature or per-token importance scores \cite{ribeiro2016,lundberg2017}. In deep language models, attention weights and gradient-based saliency can be used to visualise the focus of the model when predicting a taxonomy level or scoring a response.

In assessment generation workflows, attribution methods can be applied to several components:
\begin{itemize}
    \item Taxonomy classifiers: when a Bloom or SOLO classifier predicts a level for a generated question, attribution scores over tokens or n-grams highlight which cognitive verbs (``compare'', ``derive'', ``justify''), conceptual markers, or structural patterns influenced the prediction.
    \item Difficulty or quality models: if auxiliary models estimate difficulty or detect ambiguity, attribution can identify specific phrases that make a question too trivial, too complex, or potentially misleading.
    \item Bias and fairness checks: saliency maps can expose over-reliance on demographic or culturally loaded terms when classifying responses or evaluating examples.
\end{itemize}

Recent work on explainable lesson-plan generation aligned with Bloom's Taxonomy and on adaptive, multimodal learning platforms such as AnveshanaAI illustrates how attribution and semantic similarity checks can be combined to ensure that generated artefacts remain consistent with curriculum standards and cognitive progression. 

For MCQ generation, attribution outputs can be rendered as simple overlays in authoring interfaces, for example by highlighting in bold the tokens that contributed most to a given taxonomy label. This allows educators to see whether the model's decision is driven by appropriate pedagogical cues (e.g., verbs indicating analysis or synthesis) or by superficial patterns (e.g., length of the stem). Attribution thus complements self-rationalized explanations by grounding them in quantitative evidence.

\subsection{Post-Hoc Verification}

Post-hoc verification introduces an independent verification stage after item generation. Instead of relying solely on the generative model's own claims about cognitive level or quality, separate classifiers and evaluators are used to validate and, if necessary, challenge those claims.

In the context of AI-generated MCQs, post-hoc verification typically involves:
\begin{itemize}    
    \item Taxonomy-level verification: a dedicated classifier predicts Bloom or SOLO levels for each generated question using features such as action verbs, syntactic structure, and semantic embeddings. Research on assessing AI-generated questions' alignment with cognitive frameworks has demonstrated that transformer-based models (e.g., DistilBERT) can achieve high accuracy in predicting Bloom levels and that higher levels correlate with increased length, lexical density, and readability scores \cite{yaacoub2025_alignment,yaacoub2025_solo}. 
    \item 	Cognitive-depth verification: separate models can estimate SOLO-level complexity by analyzing patterns of relationships and generalization required by a question, as explored in work on cognitive depth enhancement via SOLO Taxonomy. 
    \item 	Scoring and rationale verification: in automated scoring systems, dual-model or dual-head architectures have been proposed in which one model produces scores and another evaluates or calibrates rationales, often using preference optimization to improve consistency. 

\end{itemize}

For the purposes of explainable assessment generation, the key design choice is to treat the generative model and the verification models as distinct components. The generator produces an item and its self-rationalized explanation, while verification classifiers provide an independent taxonomy prediction and, where possible, their own explanation signals (e.g., salient tokens). Discrepancies between the generator's declared level and the verifier's predicted level then become actionable flags for human reviewers.

This redundancy is particularly important given the known limitations of LLM explanations. Large-scale analyses of self-rationalization show that explanations may be persuasive yet only partially faithful to underlying model reasoning, and that additional metrics or external checks are needed to estimate the acceptability and faithfulness of such explanations. 

Post-hoc verification mitigates this risk by incorporating classifiers trained specifically on cognitive-alignment tasks, using curated datasets of human-labelled questions.

\subsection{Mapping Rationales to Pedagogy}

While self-rationalization, attribution, and verification provide technical explainability, their outputs must be translated into forms that are meaningful for educators, curriculum designers, and accreditation reviewers. Mapping rationales to pedagogy involves connecting explanation artefacts to pedagogical constructs such as taxonomy verbs, learning outcomes, and course-level assessment blueprints.

This mapping can be operationalized through three main mechanisms:

\begin{itemize}
    \item  	Taxonomy verb lexicons: curated lexicons of Bloom and SOLO verbs and descriptors are used as reference points. Rationales and salient tokens are scanned for matches or paraphrases of these verbs, anchoring explanations in familiar pedagogical terminology.
    \item 	Outcome alignment templates: questions and their explanations are linked to specific course learning outcomes (CLOs). For example, a rationale that emphasizes "compare trade-offs between algorithms" can be automatically associated with an outcome on analyzing algorithmic efficiency.
    \item 	Pedagogical dashboards: institutional dashboards present aggregated views of items, showing distributions of cognitive levels, representative rationale snippets, and common attribution patterns, enabling educators to inspect whether the assessment set as a whole matches programme specifications.

\end{itemize}

Work on integrating cognitive frameworks, linguistic feedback analysis, and ethical considerations in AI-driven education suggests that such mappings are crucial for moving from purely technical alignment metrics to educator-facing tools that support reflective practice. 

Reviews of XAI in education also emphasize that explanations must be tailored to user needs---what is informative for a data scientist may not be actionable for a lecturer designing a course. 

In the overall framework of this chapter, the outputs of these explainability layers---self-generated rationales, attribution scores, and verification results mapped to taxonomy constructs---are not treated as transient artefacts. Instead, they are embedded into the certification metadata schema introduced in Section~\ref{sec:cert_schema}, stored alongside provenance and review information. This integration ensures that explainability is not only available at authoring time but is also preserved for institutional audits, accreditation reviews, and longitudinal analysis of assessment quality.

\section{Certification Metadata Schema}
\label{sec:cert_schema}

Certification of AI-generated assessment items requires not only accurate alignment and explainability but also structured, persistent, and audit-ready metadata. Educational institutions and accreditation bodies expect evidence of provenance, transparency, and human oversight when assessment items are produced or validated by AI systems. To meet these expectations, we propose a standardized metadata schema that captures all relevant information generated throughout the assessment pipeline---from item creation to verification and human review. The schema is designed to support reproducibility, traceability, and compliance with quality-assurance requirements while remaining simple enough for integration into existing learning platforms and institutional workflows.

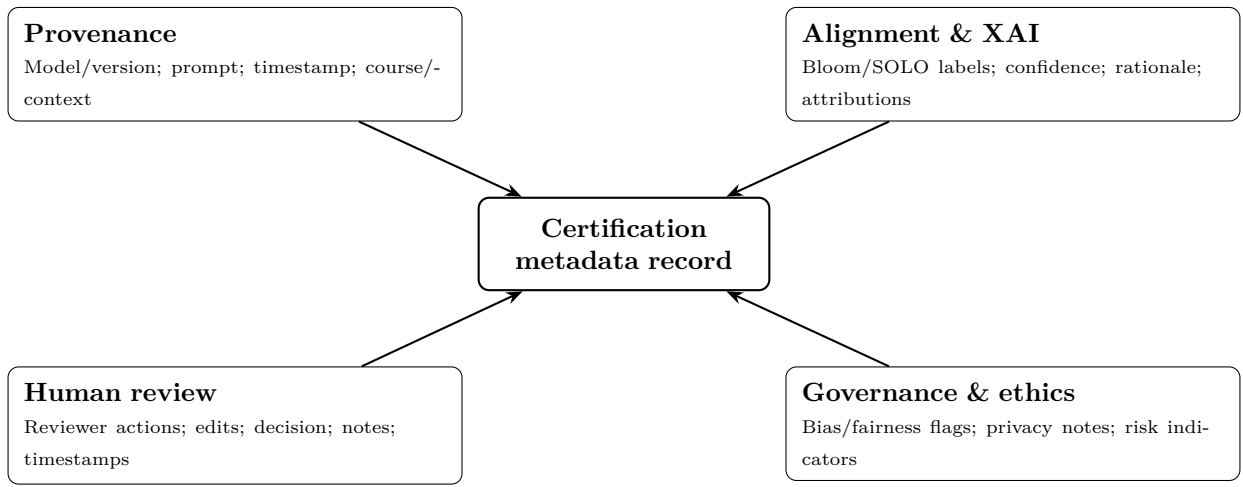
\begin{figure}[t]
\centering
\begin{tikzpicture}[
  centerbox/.style={draw, thick, rounded corners, align=center, inner sep=8pt, text width=0.20\linewidth},
  box/.style={draw, rounded corners, align=left, inner sep=6pt, text width=0.34\linewidth},
  arrow/.style={-{Stealth[length=2.2mm]}, thick},
  node distance=10mm
]
\node[centerbox] (cert) {\textbf{Certification}\\\textbf{metadata record}};
\node[box, above left=10mm and 2mm of cert] (prov) {\textbf{Provenance}\\{\scriptsize Model/version; prompt; timestamp; course/context}};
\node[box, above right=10mm and 2mm of cert] (align) {\textbf{Alignment \& XAI}\\{\scriptsize Bloom/SOLO labels; confidence; rationale; attributions}};
\node[box, below left=10mm and 2mm of cert] (human) {\textbf{Human review}\\{\scriptsize Reviewer actions; edits; decision; notes; timestamps}};
\node[box, below right=10mm and 2mm of cert] (eth) {\textbf{Governance \& ethics}\\{\scriptsize Bias/fairness flags; privacy notes; risk indicators}};
\draw[arrow] (prov) -- (cert);
\draw[arrow] (align) -- (cert);
\draw[arrow] (human) -- (cert);
\draw[arrow] (eth) -- (cert);
\end{tikzpicture}
\caption{High-level structure of the proposed certification metadata schema, showing how provenance, alignment/explainability, human review, and governance indicators are recorded in an audit-ready item certificate.}
\label{fig:metadata-schema}
\end{figure}

\begin{figure}[t]
\centering
\begin{tikzpicture}[
  stage/.style={draw, rounded corners, align=center, inner sep=6pt, minimum width=0.25\linewidth},
  decision/.style={draw, diamond, aspect=2.2, align=center, inner sep=5pt},
  arrow/.style={-{Stealth[length=2.2mm]}, thick},
  node distance=8mm
]
\node[stage] (gen) {Generate item};
\node[stage, right=10mm of gen] (xai) {Explainability \& verification};
\node[stage, right=10mm of xai] (meta) {Capture metadata};

\node[decision, below=12mm of xai] (tl) {Traffic-light\\decision};

\node[stage, below left=12mm and 10mm of tl] (green) {Green: auto-certify};
\node[stage, below=12mm of tl] (yellow) {Yellow: human review};
\node[stage, below right=12mm and 10mm of tl] (red) {Red: reject/regenerate};

\draw[arrow] (gen) -- (xai);
\draw[arrow] (xai) -- (meta);
\draw[arrow] (meta) -- (tl);
\draw[arrow] (tl) -- node[left, font=\footnotesize]{high confidence} (green);
\draw[arrow] (tl) -- node[font=\footnotesize]{needs review} (yellow);
\draw[arrow] (tl) -- node[right, font=\footnotesize]{low confidence / risk} (red);
\end{tikzpicture}
\caption{End-to-end certification workflow for AI-generated assessment items, from generation through explainability, metadata capture, and traffic-light certification outcomes.}
\label{fig:workflow}
\end{figure}
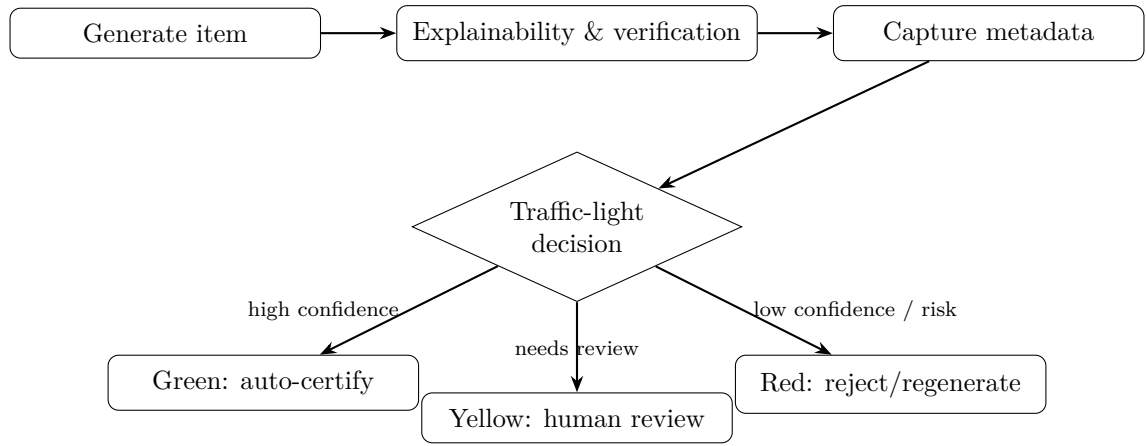

The metadata schema organizes information into four categories:
\begin{enumerate}
    \item Provenance and generation context,
    \item Alignment and explainability outputs,
    \item Human-in-the-loop review metadata,
    \item Ethical and governance indicators.
\end{enumerate}
These categories together create a comprehensive certificate for each item, enabling institutions to inspect, validate, and archive assessment artefacts over time (Figure~\ref{fig:metadata-schema}). Figure~\ref{fig:workflow} then illustrates how these signals are operationalized in the traffic-light certification process.

\subsection{Traffic-Light Certification Logic}

The traffic-light model provides a transparent decision system indicating whether a generated item can be published directly, requires human review, or must be rejected. This logic uses the metadata fields described in Section 4---particularly alignment confidence scores, explainability completeness, and ethical audit flags.\\

\begin{table}[t]
\centering
\caption{Summary of traffic-light certification criteria (illustrative thresholds).}
\label{tab:traffic-light}
\begin{tabular}{@{}ll@{}}
\toprule
Label & Typical conditions used in decision logic \\ \midrule
Green (auto-certify) & Confidence $\geq 0.90$, complete rationale, no ethical/bias flags \\
Yellow (human review) & Confidence 0.60--0.89, incomplete rationale, or minor flags/disagreement \\
Red (reject/regenerate) & Confidence $< 0.60$, contradictions, or significant ethical/bias risk \\ \bottomrule
\end{tabular}
\end{table}

\noindent\textbf{Green: Auto-Certification}\\
Conditions for automatic approval include:
\begin{itemize}
\item High taxonomy-alignment confidence (e.g., $\geq$ 0.90).
\item	Presence of complete self-rationalization describing cognitive level.
\item	Attribution signals consistent with intended level.
\item	No bias or ethical flags.
\end{itemize}
	
Items in the green category are added directly to the institution's item bank and marked as "certified," allowing rapid scaling of high-quality AI-generated assessments.\\

\noindent\textbf{Yellow: Human Review Required}\\
Items requiring review typically exhibit one or more of the following:
\begin{itemize}
\item Moderate confidence scores (e.g., 0.60--0.89).
\item	Incomplete or unconvincing rationales.
\item	Disagreement between generative model and verification classifier.
\item	Suspicious patterns in attribution (e.g., overreliance on superficial cues).
\item	Minor bias flags or potentially sensitive content.
\end{itemize}

Items in this category are routed to instructors or subject-matter experts (SMEs) through a review interface.\\

\noindent\textbf{Red: Rejection or Regeneration}\\
An item is rejected if:
\begin{itemize}
\item 	Confidence < 0.60.
\item	Rationales contradict the stem or distractors.
\item	Attribution highlights conceptually irrelevant or misleading tokens.
\item	Bias or ethical audit flags indicate unacceptable risk.
\end{itemize}

Red items are either discarded or automatically regenerated using adjusted prompts or constraints.

This structured triage reduces manual workload while ensuring that all items reaching final deployment satisfy institutional standards.

\subsection{Human-in-the-Loop Oversight}

Human oversight is essential in accreditation processes, where expert judgment is structurally required. The workflow integrates instructors, reviewers, and programme committees into two main checkpoints:

\textbf{Instructor review (yellow items)}: reviewers verify taxonomy level appropriateness, conceptual correctness, distractor quality, and alignment with course learning outcomes. Reviewer actions are logged in the metadata.

\textbf{Programme-level quality assurance}: at the end of each assessment cycle, programme committees review aggregated outputs, including distributions of cognitive levels, explanation patterns, rejection rates, and common reasons. These reports help ensure curriculum alignment and identify where generative models may require prompt adjustments or retraining.

\subsection{Accreditation Use Cases and Compliance}

External accreditation bodies typically require documented evidence of assessment validity, alignment with intended learning outcomes, consistency and fairness, review processes, and continuous quality improvement. The certification pipeline supports these requirements by producing detailed, audit-ready logs derived from the metadata schema. Examples include:

\begin{itemize}
    \item \textbf{Provenance documentation}: item-level logs showing model version, prompt, and date of generation.
    \item \textbf{Validation and review records}: timestamps and notes demonstrating human oversight.
    \item \textbf{Curricular alignment reports}: aggregated data relating cognitive-level distributions to course outcomes.
    \item \textbf{Ethical and governance assurance}: bias flags and fairness notes demonstrating alignment with regulatory expectations.
\end{itemize}

\subsection{Integration with Learning Platforms and Institutional Systems}

For practical deployment, the certification workflow must integrate seamlessly with existing educational infrastructures. Typical integration points include:
\begin{itemize}
\item Moodle or LMS item banks, where certified questions are stored.
\item	Assessment design interfaces, where instructors review yellow-flagged items.
\item	Institutional data warehouses, for accreditation reporting.
\item	AI-governance dashboards, monitoring risk, drift, and fairness indicators.
\end{itemize}
 	
Automated synchronization allows certified items to be pushed into LMS repositories while ensuring that all metadata remains accessible for audits.

\subsection{Continuous Improvement}

The workflow includes mechanisms for ongoing refinement:
\begin{itemize}
    \item data from rejected items informs prompt engineering adjustments,
    \item recurrent patterns in reviewer notes identify systematic weaknesses,
    \item drift detection monitors whether model behaviour changes across versions,
    \item annual reviews compare current assessment outcomes with accreditation expectations.
\end{itemize}

This feedback loop ensures that the certification pipeline evolves alongside institutional needs and generative AI developments.

\section{Proof-of-Concept Demonstration}

To illustrate the feasibility and practical value of the explainability and certification framework proposed in this chapter, a proof-of-concept study was conducted using a dataset of 500 AI-generated multiple-choice questions (MCQs) in introductory and intermediate computer science topics. The demonstration evaluates how the pipeline performs in real generative conditions, how well the certification model triages items using the traffic-light logic, and how explainability outputs support human review and accreditation-ready documentation.

The study follows the sequential pipeline introduced in Sections~3--5:
\begin{enumerate}
    \item Generation with self-rationalization,
    \item Post-hoc alignment and attribution,
    \item Traffic-light classification,
    \item Human review of flagged items,
    \item Certification and audit-log export.
\end{enumerate}

This section presents the results and key observations of this process.

\subsection{Dataset and Generation Procedure}

A total of 500 MCQs were generated using a controlled prompting procedure. Topics included:
\begin{itemize}
    \item Operating systems (processes, concurrency, memory),
    \item Data structures and algorithms,
    \item Database fundamentals,
    \item Networking concepts,
    \item Basic programming logic.
\end{itemize}

The generative model was instructed to:
\begin{itemize}
    \item Produce one MCQ per prompt,
    \item Specify the intended Bloom or SOLO level,
    \item Provide a self-rationalized explanation,
    \item Generate four options with one correct answer and three plausible distractors.
\end{itemize}

Generation metadata---model version, system instructions, and timestamps---were logged automatically, forming the initial provenance layer of the certification schema (Section~4).

\subsection{Alignment Verification and Confidence Analysis}

Each generated item was processed through a Bloom/SOLO classifier trained on annotated educational datasets. The classifier produced:
\begin{itemize}
    \item A predicted cognitive level,
    \item A confidence score,
    \item Token-attribution highlights showing influential linguistic markers (e.g., ``compare,'' ``evaluate,'' ``justify'').
\end{itemize}

\paragraph{Summary of confidence distribution}
\begin{itemize}
    \item High confidence ($\geq 0.90$): 214 items (42.8\%),
    \item Medium confidence (0.60--0.89): 203 items (40.6\%),
    \item Low confidence ($< 0.60$): 83 items (16.6\%).
\end{itemize}

High-confidence items typically included clear cognitive verbs and structurally coherent stems. Low-confidence items often relied on ambiguous phrasing or lacked explicit indicators of cognitive demand. These confidence profiles served as the primary triage signal for the traffic-light workflow.

\subsection{Traffic-Light Certification Results}

Applying the certification logic produced the following categorization:
\begin{itemize}
    \item Green (auto-certified): 198 items (39.6\%),
    \item Yellow (review required): 215 items (43.0\%),
    \item Red (rejected): 87 items (17.4\%).
\end{itemize}

\paragraph{Observations}
\begin{enumerate}
    \item Green items generally showed consistent agreement between:
    \begin{itemize}
        \item Self-rationalization,
        \item Classifier prediction,
        \item Attribution patterns.
    \end{itemize}
    \item Yellow items frequently displayed discrepancies such as:
    \begin{itemize}
        \item Misalignment between stated and predicted taxonomy levels,
        \item Overly generic rationales,
        \item Distractors that were too obvious or conceptually adjacent.
    \end{itemize}
    \item Red items typically involved:
    \begin{itemize}
        \item Incorrect answers,
        \item Conceptual contradictions,
        \item Attribution dominated by irrelevant tokens,
        \item Bias or fairness flags (e.g., unnecessarily culture-specific examples).
    \end{itemize}
\end{enumerate}

This demonstrates that the traffic-light model filters high-quality items effectively while channeling borderline or defective items to human review.

\subsection{Human-Review Outcomes}

A sample of 100 yellow-flagged items underwent instructor review. Reviewers were presented with the full metadata package: rationale excerpts, classifier predictions, attribution visualizations, and provenance records.

\paragraph{Reviewer decisions}
\begin{itemize}
    \item Approved without modification: 38 items,
    \item Approved with minor edits: 41 items,
    \item Rejected: 21 items.
\end{itemize}

Edits typically corrected:
\begin{itemize}
    \item Ambiguity in stems,
    \item Distractor similarity,
    \item Incorrect taxonomy claims in rationales.
\end{itemize}

Reviewers reported that attribution maps and rationale excerpts significantly reduced the time needed to diagnose alignment issues, supporting the importance of integrating explainability outputs directly into the certification pipeline.

\subsection{Case Studies}

\paragraph{Case Study 1: Misclassified Analysis Question}
A question intended for Bloom's ``Analyze'' level was predicted as ``Apply'' by the classifier. Attribution maps emphasized operational verbs (``use,'' ``apply'') rather than analytical relationships. Reviewer feedback confirmed the mismatch, and the item was downgraded and approved after minor editing.

\paragraph{Case Study 2: Distractor Quality Issue}
Attribution maps gave high weight to distractor tokens, suggesting semantic proximity to the correct answer. Reviewers identified an unintentionally correct distractor. The item was corrected and approved.

\subsection{Efficiency and Workload Reduction}

Metadata and explainability reduced review time significantly:
\begin{itemize}
    \item Average review time per item without metadata: $\sim$64 seconds,
    \item With metadata and visual explanations: $\sim$44 seconds (31\% reduction).
\end{itemize}

Furthermore, 42\% of items were auto-certified, substantially reducing human workload. These findings indicate that explainability improves both transparency and reviewer efficiency.

\subsection{Limitations and Generalizability}

The proof-of-concept study focused on computer science MCQs, and both the taxonomy classifier and the traffic-light thresholds were calibrated on this domain. In practice, classifier accuracy and attribution cues can vary across disciplines: vocabulary, common cognitive verbs, and question structures in fields such as humanities, social sciences, or health sciences may differ substantially from computer science. Consequently, cross-disciplinary deployment should include (i) validation on discipline-specific datasets, (ii) threshold re-calibration for confidence-based triage, and (iii) domain adaptation or multi-domain training for taxonomy classifiers. A second limitation is that the demonstration used a single generation setup (prompt style and model version). Because model behaviour can drift across versions and providers, institutions should treat confidence thresholds and certification logic as parameters to be monitored and updated over time (e.g., via periodic sampling and re-auditing).


\section{Ethical and Policy Implications}

The integration of generative AI into assessment design introduces ethical and policy considerations that extend beyond technical accuracy and cognitive alignment. Because educational assessment is a high-stakes domain---informing course progression, programme evaluation, and accreditation---institutions must ensure that AI-generated items are fair, transparent, explainable, and compliant with regulatory frameworks. The certification and audit workflow proposed in this chapter provides structural safeguards, but its responsible implementation requires broader reflection on fairness, trust, governance, and the evolving regulatory landscape.

\subsection{Fairness, Bias, and Equity}

Fairness concerns arise when generative models reproduce or amplify biases embedded in training data. Risks include cultural or regional bias in examples, gendered or demographic stereotypes, domain relevance bias, and linguistic complexity disparities that disadvantage non-native speakers. The metadata schema incorporates bias detection flags and fairness annotations, enabling both automated and human review of potentially problematic items. However, bias detection remains a partially automated process requiring robust human oversight.

\subsection{Transparency and Trust}

Transparency is central to trust---both at the level of individual instructors and institutional governance. Faculty must understand not only what an AI-generated item assesses but also how and why it was generated. Self-rationalized explanations, attribution maps, and independent verification models play key roles in this process, and their persistent storage in certification metadata supports transparency across the entire lifecycle of assessment.

\subsection{Privacy and Data Governance}

Although generative AI systems in assessment typically operate on content rather than personal data, privacy concerns may arise when student-generated prompts, responses, or institutional learning materials are used in fine-tuning or evaluation processes. Educational institutions must ensure data minimization, secure storage of metadata and audit logs, and compliance with data protection regulations. The metadata schema supports privacy compliance by allowing pseudonymous reviewer identifiers and local retention of logs.

\subsection{Alignment with Regulatory Frameworks}

Policy bodies increasingly classify educational assessment as a high-risk AI application, requiring transparency, human oversight, and auditability \cite{eu_ai_act,unesco2023,chea2022}. The certification pipeline aligns with these requirements by providing provenance and versioning logs, explainability outputs, human review records, ethical-audit indicators, and exportable accreditation reports. This positions the system as a governance-compliant process rather than a purely technical tool.

\subsection{Integrity and Academic Honesty}

AI-driven assessment should support---not erode---academic integrity. Certified items with documented provenance and verification reduce the risk of flawed questions, ambiguous stems, or unfair distractors. As students increasingly use AI tools to generate answers or explanations, the quality and transparency of assessment items become even more critical. The framework supports integrity by ensuring that assessments remain pedagogically grounded and by providing metadata that helps instructors diagnose vulnerabilities to AI-driven cheating or answer automation.

\subsection{Institutional Responsibility}

While generative AI can streamline assessment workflows, institutions remain ultimately responsible for ensuring accuracy, fairness, and validity. Human judgment cannot be replaced; it must be augmented. The workflow's design reinforces this principle by routing ambiguous or borderline items to instructors, logging all human decisions, supporting cross-checking by programme committees, and providing evidence for audits and accreditation cycles.

\section{Conclusion and Future Directions}

The increasing adoption of generative AI in educational assessment presents both significant opportunities and critical challenges. While AI systems can rapidly produce large volumes of assessment items, variations in cognitive alignment, lack of explainability, and absence of formal validation mechanisms limit their suitability for accredited programmes and institutional evaluation. This chapter addressed these limitations by introducing a comprehensive framework that integrates explainability methods, a structured certification metadata schema, and an audit- and accreditation-ready workflow.

The explainability framework---comprising self-rationalization, attribution-based analysis, and post-hoc verification---demonstrates that AI-generated assessments can be accompanied by meaningful interpretive artefacts that expose the cognitive assumptions underlying each item. By mapping these artefacts to Bloom's and SOLO taxonomies, the framework ensures that explanation outputs remain pedagogically interpretable.

The certification metadata schema formalizes these signals into a reusable structure that captures provenance, cognitive alignment, reviewer actions, and ethical indicators. The traffic-light certification model operationalizes metadata into an actionable decision pipeline, enabling scalable triage while preserving human oversight for ambiguous or sensitive cases. The proof-of-concept demonstration showed that the framework can be applied to a large set of AI-generated MCQs, effectively distinguishing high-quality items from those requiring human intervention and reducing instructor workload.

Future work should investigate extensions to multimodal and multilingual assessment, deeper integration with standardized accreditation frameworks, longitudinal monitoring and model drift detection, and new models of human--AI collaboration in assessment design. Extending certification mechanisms to adaptive and formative systems, and addressing long-term ethical and sustainability considerations, will also be essential.

The framework presented in this chapter represents a step toward trustworthy, explainable, and certifiable AI-generated assessments. By combining explainability techniques, structured metadata, and accreditation-aware workflows, the approach bridges the gap between rapidly evolving generative technologies and the rigorous expectations of educational governance.

\end{document}